\documentclass[aps,prb,amsmath,amssymb,twocolumn,showpacs,floatfix,reprint,superscriptaddress,nobibnotes, longbibliography]{revtex4-1} 
\usepackage[utf8]{inputenc} 
\usepackage{graphicx}
\usepackage{dcolumn}
\usepackage{bm}
\usepackage{units}
\usepackage{amsmath}
\usepackage{times}
\usepackage[caption=false]{subfig}
\usepackage{xcolor}
\graphicspath{{./figures/}}

\begin{document}
	\newcommand{\bb}{\textbf{b$_0$} }
	\newcommand{\bbb}{\textbf{b$_0$}}
	\newcommand{\cc}{\textbf{c$_0$} }
	\newcommand{\nbse}{NbSe$_3$}
	\newcommand{\comment}{\textcolor{blue}}
	\newcommand{\todo}{\textcolor{red}}
	\newcommand{\q}{\textbf{q$_1$} }
	\newcommand{\qe}{\textbf{q$_1$}}
	\newcommand{\qs}{\textbf{q$_1$'} }
	\newcommand{\qb}{q$_{1b}$ }
	\newcommand{\qq}{\textbf{q$_2$} }
	\newcommand{\qqe}{\textbf{q$_2$}}
	\newcommand{\qqs}{\textbf{q$_2$'} }			
	\newcommand{\qqb}{q$_{2b}$ }
	\newcommand{\T}{T$_1$}
	\newcommand{\TT}{T$_2$}

\title{Spatial ordering of the charge density waves in NbSe$_3$}
\author{M. A. van Midden}
\author{H. J. P. van Midden}
\author{A. Prodan}
\affiliation{Jožef Stefan Institute, Jamova 39, SI-1000 Ljubljana, Slovenia}

\author{J. C. Bennett}
\affiliation{Department of Physics, Acadia University, Wolfville, Nova Scotia, Canada B4P 2R6}

\author{E. Zupanič}
\affiliation{Jožef Stefan Institute, Jamova 39, SI-1000 Ljubljana, Slovenia}

\begin{abstract}
Ordering of the two incommensurate charge density waves (CDW), \mbox{\q = (0.0, 0.243, 0.0)} and \mbox{\qq = (0.5, 0.263,0.5)} in the quasi-one-dimensional \nbse ~structure is studied by means of low temperature scanning tunneling microscopy. Larger (100) van der Waals surfaces are analyzed using one-dimensional Fourier transforms along the three types of trigonal prismatic columns, running parallel to the monoclinic \bb direction. The procedure enables unambiguous differentiation between both CDWs modulating individual columns. In addition, it allows a quantitative comparison of modulation amplitudes along different columns of the same type and, in case of sufficiently large images, also along individual columns. 
It is confirmed that each CDW modulates both type-III and type-I columns with the narrowest isosceles bases. As a consequence beating is observed between the two slightly different modes. The bridging type-II columns appear not to be modulated. On a large scale the modulation amplitudes along individual columns vary, suggesting formation of CDW nanodomains. The possibility of interchanging both CDWs along columns forming symmetry-related pairs results in a charge difference, which might be the possible origin of CDW sliding.
\end{abstract}

\maketitle

\section{Introduction}

\begin{figure}[t!]
	\begin{center}
		\includegraphics[width=1
		\linewidth]{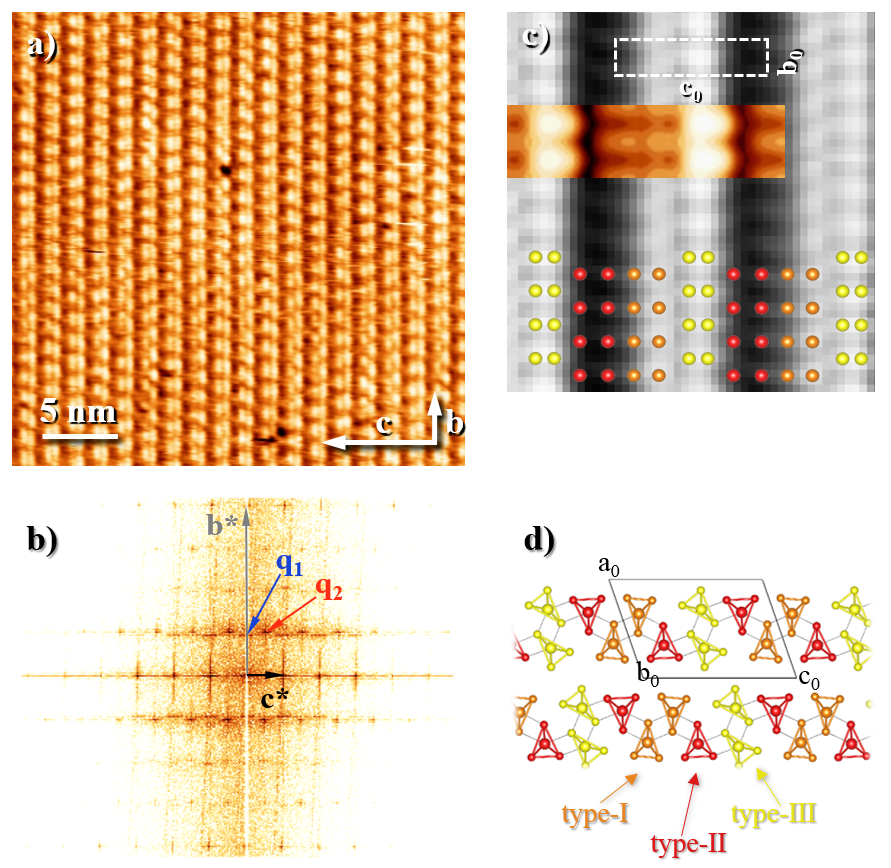}
		\caption{a) Constant current LT STM image of a 30nm x 30nm large (100) van der Waals surface of \nbse ~\textit{(scanning parameters: \mbox{V$_{bias}$ = -470 mV}, I = 100 pA)} with resolved surface Se atoms and both CDW modes. b) FT of the image in a) showing both, the basic structure reflections and the IC CDW pairs of satellites. c) Inverse FT of a part of the image shown in a) with both CDW modes removed by filtering the FT. The surface unit cell is marked with a white dashed rectangle and the coloured inset shows a DFT simulated STM image. In the lower part of the image different columns are identified by marking the positions of the uppermost Se atoms with the same colours as in d). d) A model of the \nbse ~ structure with type-I (orange), type-II (red) and type-III (yellow) bi-capped TP columns along the \bb axis.}
		\label{structure}
	\end{center}
\end{figure}

Niobium triselenide (\nbse) is one of the few quasi-one-dimensional trichalcogenides that undergo two Peierls transitions \cite{Wilson, PRODAN_2010}. It also exhibits interesting transport properties, in particular sliding of its two charge density waves (CDW) in an external electric field \cite{Monceau+Ong_1977}. As a result it was thoroughly studied in the past by a variety of different techniques from X-ray diffraction \cite{Wilson, Xray_q2_1996, vansmaalen}, Selected Area Electron Diffraction (SADF) Transmission Electron Microscopy (TEM) \cite{Fung-Steeds_1980}, Angle Resolved Photoemission Spectroscopy (ARPES) \cite{ARPES_arxiv,ARPES_PRL_2017,ARPES_PRB_2019} as well as Scanning Tunneling Microscopy (STM) \cite{Monceau_Electronic-crystals-overview,BRUN2015_PhysB,Brun2009_PRB,Brun2012_PhysB,Brun_PRL_2010,Soliton_PRL2012}.
Each of the CDW modes, with \mbox{\q = (0.0, 0.243, 0.0)} and \mbox{\qq = (0.5, 0.263, 0.5)} with onset temperatures T$_1$ = 144 K and T$_2$ = 59 K, were originally suggested \cite{Wilson} to selectively modulate one of the three available types of trigonal-prismatic (TP) columns, which form the monoclinic unit cell of the \nbse~basic structure (\mbox{a$_0$ = 10.009 \AA}, \mbox{b$_0$ = 3.480 \AA}, \mbox{c$_0$ = 15.629 \AA}, \mbox{$\beta_0$ = 109.47$^{\circ}$}, space group P2$_1$/m)\cite{Wilson,Brun2009_PRB}. These two column types with the narrowest equilateral bases are usually described as type-III or "yellow" and type-I or "orange", while the remaining "more isosceles-like" type-II or "red"  columns were supposed to remain unmodulated. A projection of the room temperature (RT) basic structure along the bi-capped 8-coordinated TP columns is presented in Fig. \ref{structure} (d).  The adjacent out-of-phase columns are inter-connected with strong covalent Nb-Se bonds and form layers, separated by van der Waals (vdW) gaps. Therefore the structure and the related properties of \nbse ~are highly anisotropic, showing both one-dimensional (1D) (along the \bb direction) and two-dimensional (2D) (parallel to the (100) vdW planes) characters.
The original model with the two CDW modes modulating the type-III and type-I columns \cite{Wilson} was in accord with a series of experiments. However, it does not explain the observed unique transport properties, which differentiate \nbse ~(and the few structurally related compounds) from the rest of the quasi-1D family of structures \cite{PRODAN_2010,Alternative-models-of-modulation,Prodan_modulated_low_dim_syst}. Doubts regarding the model were supported after a careful re-examination \cite{Prodan_modulated_low_dim_syst} of the first published LT STM image of \nbse \cite{Coleman,Comment_van-Smaalen,Coleman_reply} and after some details were clearly revealed in the recently published LT STM experiments \cite{MONCEAU2015_modulation2D,BRUN2015_PhysB,Brun2009_PRB,Brun2012_PhysB,Brun_PRL_2010}. 

A proper understanding of the way CDWs slide in the few quasi-1D compounds is of crucial importance, because if as originally suggested, it indeed represents a novel conduction mechanism. The fact that LT STM results are the only ones not in full accord with the previously suggested modulated structure of \nbse \cite{Wilson} cannot be without reasons. It may be argued that the discrepancy should be attributed to STM being only sensitive to the sample surface. However, this argument cannot be unconditionally accepted, particularly not in case of layered materials, where a relatively strong contribution from subsurface atomic layers was often detected under proper scanning conditions. Another important difference with regard to all other experimental methods is the fact that STM is a local method that enables studying the sample surface not only with atomic resolution, but also on areas sufficiently small to avoid averaging, which may in special cases lead to false conclusions \cite{PRODAN_2010,Prodan_modulated_low_dim_syst,Alternative-models-of-modulation}. 

In the present work detailed analysis of large-scale LT STM images of the (100) vdW surfaces of \nbse , recorded under different tunneling conditions, are presented. The examined images reveal atomic resolution in addition to clearly resolved CDW modes. Careful 1D Fourier transform (FT) analysis gives exact connection between both modes and the column types they modulate. In addition, it enables studies of variations in intensity of the two modes between different (as well as along the same) columns of the quasi-1D structure.

\section{Experimental}
All measurements were performed using a Joule-Thomson STM at 4.2 K or 1.2 K, i.e. well below both CDW transition temperatures. All images presented were recorded in constant current mode, using an etched tungsten tip cleaned and preliminary verified on Cu(111) single crystal surfaces. Bias voltages refer to the
sample voltage with respect to the tip. 

The \nbse ~ crystals were grown by physical vapour transport within a three-zone furnace.  Stoichiometric amounts of Nb (99.999) and Se powders (99.98) were sealed in an evacuated quartz ampule along with a few milligrams of iodine as a transport agent.  The ampule was slowly heated with a reverse gradient to 670 – 630 $^\circ$C to react the powder.  A forward gradient of 670 - 700 $^\circ$C was then established and maintained for a period of 10 days, after which the ampule was cooled to 550 $^\circ$C over 10 hours and quenched to RT. 

The samples were cleaved along the vdW (100) planes \textit{in situ} in ultra-high vacuum (UHV) at RT, just before they were transferred into the measuring head and cooled to the measuring temperature. 

Density functional theory (DFT) calculations were performed using the Quantum-ESPRESSO (QE) code \cite{QE-DFT}. The electron-ion interactions are described by the Garrity-Bennett-Rabe-Vanderbilt (GBRV) high throughput ultrasoft pseudopotentials \cite{DFT}, with a set of highly transferable pseudopotentials, able to reproduce metallic, ionic, and covalent bonding behaviors. The procedure consistently delivered accurate results for all involved atoms in a variety of crystal structures. The GBRV pseudopotentials utilize the PBE exchange–correlation functional \cite{vdW-DFT} and fixed plane-wave values of 40 Ry and charge density cutoffs of 200 Ry.
	
A (2,2,2) Monckhorst-Pack k-point sampling grid was used to calculate the structural properties of the bulk basic structure of \nbse. The lattice parameters and the atomic structure were relaxed until the changes in total energy were below $1.0 \cdot 10^{-6}$ Ry and the forces below $1.0 \cdot 10^{-5}$ atomic units. One layer of \nbse ~and 5 layers of vacuum were once again relaxed with the given lattice parameters in an orthogonal unit cell providing the basic structure for STM simulations, which were performed using the QE package utilizing the Tersoff-Hamann \cite{Tersoff-Hamann} approach. The final DFT parameters of the \nbse ~basic structure are a = 10.615\AA, b = 3.6602 \AA, c = 15.604 \AA, $\beta$ = 109.16$^\circ$.

\section{Results}

\subsection{Identification of the surface unit cell and the columns}

In order to study the modulated structure, the surface unit cell and the TP columns had to be identified first. Based on the known RT structure \cite{vansmaalen} presented in Fig. \ref{structure} d) the direction of the TP columns can be unambiguously determined in Fig. \ref{structure} a). Only part of the constant current image is presented for clarity. The modulation along particular columns is also clearly resolved. The FT of the entire image reveals the surface unit cell of the basic structure and the \q mode with \qb= 0.246 b* as well as the enlarged unit cell of the \qq mode with \mbox{\qqs = (0.26 b*, 0.5 c*)}, all in accord with previous publications. In Fig. \ref{structure} c) a smaller part of the inverse FT image with contributions of both CDW modes omitted is presented, clearly showing a surface unit cell consisting of three distinct TP columns of which one is shifted by half a unit cell along \bbb. Based on the performed DFT simulations the largest contribution to the STM image comes from the top selenium atoms. Therefore, the shifted columns can be unambiguously identified as type-III or "yellow". The remaining two column types are also clearly in accord with the simulated DFT image, which is displayed in colour on top of the grayscale STM image. The described identification of the columns is also in accord with previously published STM images, where the columns that appeared the lowest at the applied bias voltage were identified as type-I or "orange" and the remaining column as type-II or "red". The top Se atoms belonging to the different TP columns are accordingly shown with balls of the same colours. 

\begin{figure}[ht!]
	\begin{center}
		\includegraphics[width=1\linewidth]{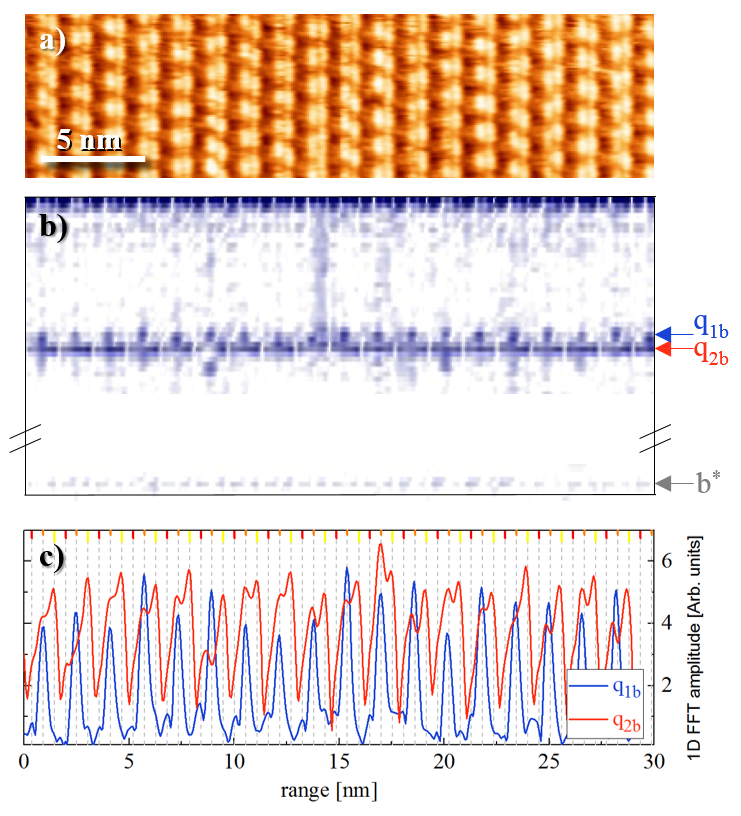}
		\caption{a) Part of the analyzed constant current LT STM image. b)1D FT performed along individual lines parallel to \bb. The gray arrow marks the position of the b* while the blue and red arrows mark the positions of the \qb and \qqb , respectively. c) Plots of 1D FT amplitudes at the marked values together with column positions obtained from a). }
		\label{short}
	\end{center}
\end{figure}

\subsection{Fourier transform analysis}

Having unambiguously identified all three TP columns the spatial distributions of both CDW modes was determined using FT analysis. Due to the limited resolution in the \textbf{k}-space, which is a result of finite image sizes, inverse 2D FTs of the contributions from either of the two modes can only provide a rough picture of which mode is present on a certain column. To avoid this problem 1D FTs were performed along individual columns, parallel to the \bb direction, taking advantage of the 1D nature of the modulated structure. During this procedure the STM image was first rotated until the \bb direction was exactly aligned parallel to the image edge. 1D FTs were calculated along all lines parallel to the edge and plotted as a function of distance perpendicular to the columns. The resulting diagram is presented in Fig. \ref{short} b) with the CDW modes marked with the blue \q and red \qq arrows. The upper panel a) presents the corresponding part of the STM image. 

Plotting the values of the 1D FTs at the marked \textbf{q}-vectors as a function of distance perpendicular to \bb gives an insight into the intensity variation of both \q and \qq modes between different columns of the same type. The positions of the TP columns are marked with ticks of the corresponding colours while dashed lines are added guides to follow their exact positions with regard to the slightly displaced CDW maxima. The analysis of the STM image, recorded at -470 mV (Fig. \ref{short} a), confirms that on the surface shown \q is mainly present on type-I columns. The \qq exhibits comparable intensities both on type-I and type-III columns. In addition the amplitudes of both modes as well as their ratios vary significantly from column to column of the same type.

\subsection{Bias dependence}

\begin{figure}[ht!]
	\begin{center}
		\includegraphics[width=0.95\linewidth]{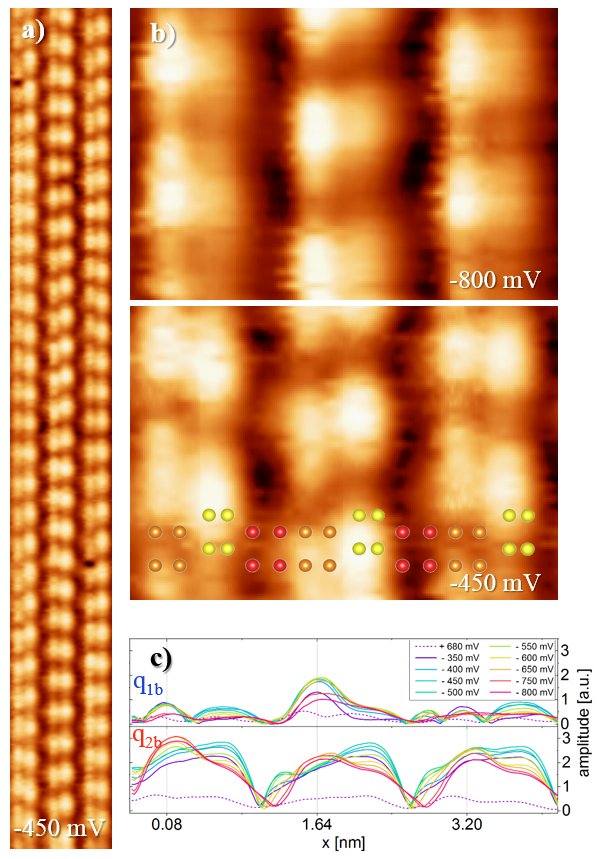}
		\caption{a) An approximately 4.7 nm $\times$ 37.5 nm area of the \nbse ~ (100) vdW surface recorded with V$_{bias}$ = -450 mV. b) and c) show 4.7 nm $\times$ 3.2 nm parts of the larger area shown in a) recorded with V$_{bias}$ = -800 mV and -450 mV, respectively. d) Plots of 1D FT amplitudes for \q and \qq modes as analyzed from images recorded at different bias voltages.}
		\label{bias}
	\end{center}
\end{figure}

The observed intensities of both \q and \qq modes in the STM images strongly depend on the bias voltages used during scanning. Fig. \ref{bias} a) shows a three unit cells wide and 37.5 nm long section, recorded at V$_{bias}$ = -450 mV, where CDW modes modulating neighbouring type-I and type-III columns show varying phase shifts.

The modulation along each column is a result of beating between \q and \qq periodicities confined to the two columns forming a pair. In Fig. \ref{bias} b) a small area recorded with bias voltages of -800 mV (top) and -450 mV (bottom) are shown for comparison. The topmost Se atoms are overlaid onto the bottom image. The 1D FT analyses (Fig. \ref{bias} c) of such images recorded at different conditions reveal the dependence of the CDW amplitudes on the bias voltage. The \q mode is dominantly observed on type-I columns at voltages down to -400 mV, while between -400 and -650 mV this mode is equally strong on both, type-I and type-III columns. At even lower voltages the \q mode is hardly observable. The amplitudes of the \qq mode are generally higher than those of the \q mode and at voltages down to -500 mV higher on type-III columns. At lower bias voltages (bellow -500 mV) the amplitudes of the \q mode on type-I columns become higher and on type-III lower. Type-II columns appear not to be modulated at all bias voltages used. There is a very weak \q and \qq modulation, best observed in STM images recorded at -500 mV, which is clearly displaced from both type-I and type-II columns. Additional careful analysis shows that this weak contribution does not represent a modulation of the interconnecting type-II columns, but rather a contribution of the modulated subsurface type-I columns. 

\begin{figure*}[ht!]
	\begin{center}
		\includegraphics[width=1\linewidth]{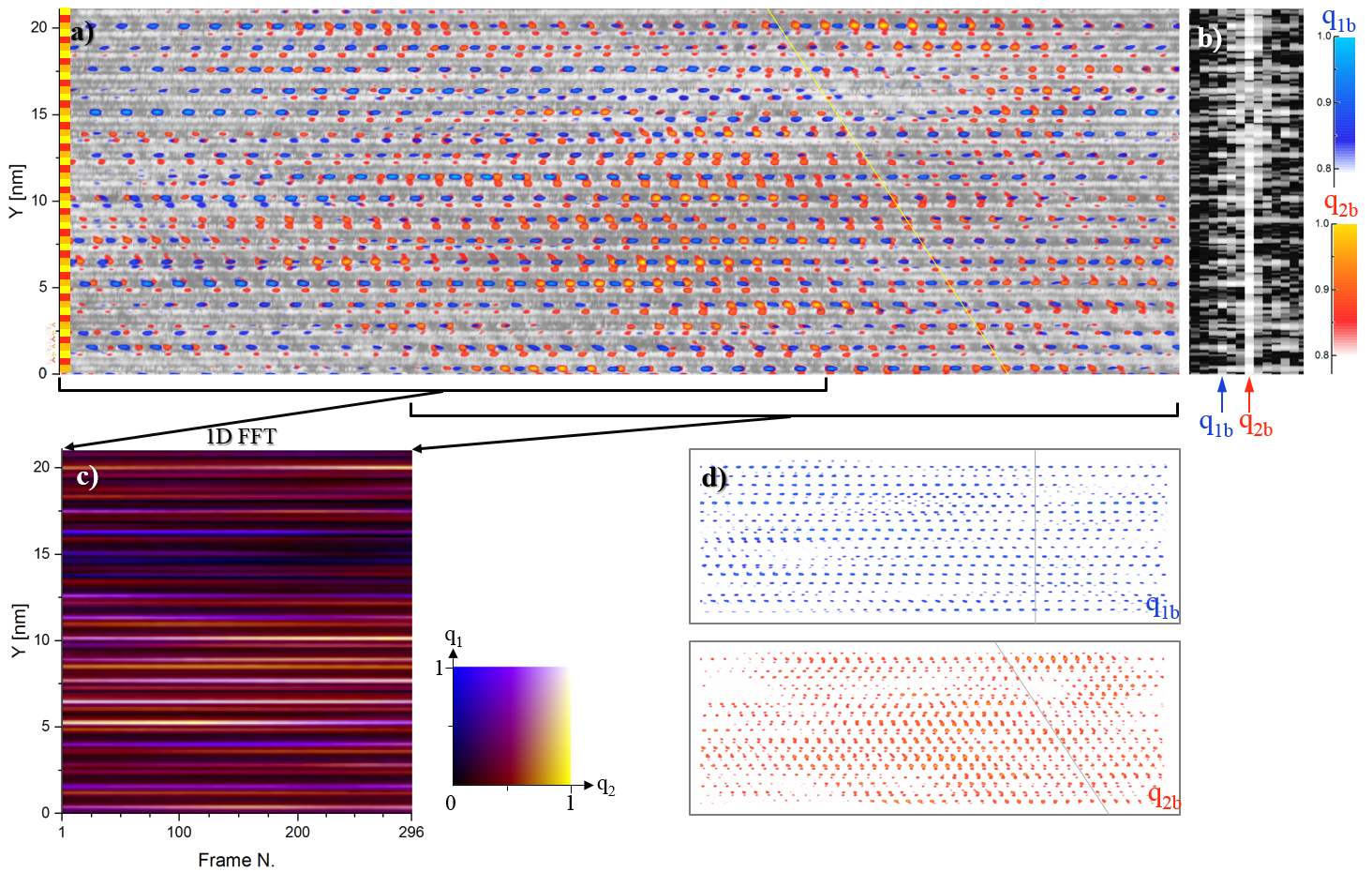}
		\caption{a) Coloured maxima of both CDW modes, extracted from the inverse 1D FTs shown in b) along the columns, properly positioned on top of the corresponding grayscale constant current STM image. c) A colour plot of the ratios of normalized amplitudes of both CDW modes (see text for details). The y-axis corresponds to the y-axis of the STM image, while the x-axis indicates the position of the window. The first and the last window positions are indicated below the image in a). d) Inverse 1D FT showing separately the \q (top) and \qq (bottom) modulation. Note the in-phase and out-of-phase relationship between CDW modulations along neighbouring columns. Two grey lines are plotted as guides.}
		\label{long}
	\end{center}
\end{figure*}

\subsection{Variations along the columns}

To study intensity variations along individual columns and variations over large surface areas a sufficiently long image along \bb is needed. Impurities that are occasionally present on the sample surface make it as a rule difficult to obtain such images with clearly resolved surface atoms on sufficiently large areas. Such an image of a 21 nm wide and 64 nm long area is shown in the background of Fig. \ref{long} a). Combined with the 1D FT it clearly reveals the presence of both modes on particular columns, shown on Fig. \ref{long} b). Using inverse FT along individual lines contributions belonging to \q and \qq CDW modes are plotted for values exceeding 80\% of their corresponding maxima with blue and red
, respectively. Plotting \q and \qq separately (Fig. \ref{long} d)) clearly reveals the in-phase relationship between \qq modes along neighbouring columns of the same type. No such clear relationship can be observed for the \q mode.

Additionally, using a 40 nm wide window (i.e. a window sufficiently wide to clearly resolve both CDW modes in the corresponding FT) 1D FT was performed on parallel lines along this limited part of the image. The amplitudes of both modes were determined for every position of this 40 nm window. Each amplitude was normalized to its \textbf{q}-vector's maximum value and their combinations were plotted using the following colour scheme: the normalized amplitudes of \qq were plotted in shades ranging from black through red to yellow, while the normalized amplitudes of \q were plotted in colours ranging from black (no modulation) to blue. If both modes appeared overlapped, the colour of the pixels was a mixture of both. Fig. \ref{long} c) was constructed by moving the window along the large STM image, performing a FT analysis at each step and plotting the corresponding vertical lines as a function of the distance (indexed as Frame N.). The performed analysis clearly reveals relatively large variations in the ratios of normalized \qe/\qq amplitudes along certain columns.

\section{Discussion}
STM and related scanning probe methods are the only ones not based on averaging. If performed under ultra clean conditions and at sufficiently low temperatures, they are in fact the only methods capable of revealing the actual local behavior of the CDWs at the atomic scale. The present STM experiments on \nbse ~were deliberately performed at somewhat lower magnifications, ensuring thus a better transparency of the CDW ordering and still revealing the required atomic resolution.

Because of its nonlinear transport properties, attributed to sliding CDWs, \nbse ~ became one of the most thoroughly studied inorganic compounds \cite{Ong_PRB_1983, Ayari_PRL_2004,Monceau+Ong_1977,Withers_1986, Gill-Wills_1986, Schaefer_2001, Hor_2005} \cite{Gruner_1983_Shapiro, book_vanSmaalen}. However, the sliding ability of the CDWs is still not properly understood. It was so far observed only in a few quasi-1D compounds, most of them structurally closely related to \nbse, e.g. the monoclinic polymorphs m-TaS$_3$ \cite{Thompson_1981} and NbS$_3$-II \cite{Zettl_1982, Zettl_PRB_1982}. A proper understanding of the phenomenon is important, because in case sliding was confirmed beyond any doubt, it would indeed represent a novel transport mechanism. It must be clarified why the CDWs can participate in the conductivity of the few cases only, while in most other quasi-1D IC compounds they can under no condition be forced into sliding. 
According to the original Wilson’s suggestion \cite{Wilson} the two IC modes in \nbse ~ are supposed to be selectively confined to the type-III (\qe) and type-I columns (\qqe), while in case of the alternative picture, based on the existence of nanodomains \cite{Alternative-models-of-modulation,Prodan_modulated_low_dim_syst}, both modes should be present along both, type-I and type-III columns. In the first case CDWs slide along TP columns after being decoupled in the presence of an external electric field from the underlying basic structure, while in the second the enhanced current is supposed to be a result of an easy switching of the two modes, which causes a displacement of the charge difference between the two modes. In both cases the consequence will be a certain CDW current. The two models can only be differentiated by determining how the two CDW modes are formed and ordered along particular TP columns. 
Although impurities can affect sliding to a certain extend, they certainly cannot be the determining factor, which will differentiate between the two categories of 1D compounds, i.e. those which always show CDW sliding and the ones that will under no circumstances be forced into it. There are in fact two important structural properties, which certainly play a decisive role in this distinction. First, \nbse ~(and likewise the related compounds) is composed of pairs of symmetry-related TP columns and second, these columns are modulated by (at least) two IC CDW modes  \cite{Prodan_modulated_low_dim_syst, Alternative-models-of-modulation}.
 The two requirements and not the concentration of the impurities appear to be the key condition for the CDWs to start sliding. 
The widely accepted belief that the two modes appear selectively along two of the three available column types in \nbse, i.e. \q = (0,0.241,0) below T$_1$ = 144 K along type-III (yellow) and \qq = (0.5,0.260,0.5) below T$_2$ = 59 K along type-I (orange) columns, was in fact seriously questioned only after some details, clearly revealed in the published LT STM images of \nbse \cite{Coleman,Comment_van-Smaalen,Coleman_reply,MONCEAU2015_modulation2D,BRUN2015_PhysB,Brun2009_PRB,Brun2012_PhysB,Brun_PRL_2010,Monceau_Electronic-crystals-overview} were re-examined. It appears that these published images are indeed in a better accord with the domain model. In fact, the older published images show similar details as the ones presented in the present work, i. e. two types of columns modulated with either of the two IC CDW modes \cite{Alternative-models-of-modulation,Prodan_modulated_low_dim_syst}. Domains can only be formed if the same CDW mode would modulate the top type-III and type-I columns. However, this mode can be likewise \q and \qqe, which necessarily leads to the formation of layered CDW domains, which differ only in the interchanged CDW modes that modulate pairs of TP columns forming a single Se-Nb-Se layer.

The intensity variation of the two modes, perpendicular to the TP columns as well as along particular columns, is in accord with the appearance of layered nanodomains. Their existence explains why CDW sliding is possible only in the few structurally related compounds with symmetry-related pairs of TP columns, both likewise able to accommodate both IC CDWs. Their commensurate combination appears to be the most energy-efficient solution for the LT ground state. 
Both models are in good accord with most experimental results, except those obtained from LT STM, which reveal very important details \cite{Brun2009_PRB,BRUN2015_PhysB,Brun2012_PhysB,Brun_PRL_2010,Coleman}. 
In particular, the LT STM images recorded at 77 K \cite{Brun2009_PRB} show in addition to the strong \q modes also weak \qq contributions, whose modulation unit cells are definitely doubled. This is either because the modes along identical columns belonging to the adjacent basic structure unit cells alternate in intensity along the \cc direction or because they appear out-of-phase. This is also a clear proof that both modes are indeed present well above \TT ~and that the "twinkling" fringes, observed in SADF TEM images\cite{Fung-Steeds_1980,Luth_surface-book} are indeed caused by the presence of both slightly different modes between both CDW onset temperatures, determined as T$_1$ and T$_2$. This is also in accord with the recently published ARPES measurements. These show that both CDW gaps persist well above their onset temperatures and  that gradual loss of long-range phase coherence takes place with rising temperature  rather than the reduction of the CDW amplitudes \cite{ARPES_arxiv}.

The intensity variation of both CDWs between and along the columns, their appearance on two types of TP columns, as well as the detection of the \qq mode well above its determined onset temperature, are obviously a result of an unstable disorder, confined to areas, smaller or at least comparable to the experimental coherence lengths in diffraction experiments.

It should finally be noted that all mentioned details, observed and confirmed in a series of LT STM studies, suggest that a structural x-ray refinement should, rather than with both CDW modes confined selectively to the type-I and type-III TP columns, be performed with both types of columns modulated by a single, highly anharmonic commensurate mode of the form \q + \qqe
. The final R-factors might in that case be reduced in comparison with those, obtained with the two modes confined selectively to the two types of columns. Such a combined mode of no physical meaning, would better mimic the disorder between both CDW modes at or even below the nanometer scale. 

\section{Conclusions}
It is demonstrated that LT STM in combination with 1D FT analysis is a powerful method for studying the CDW properties in low-dimensional structures. In the presented case of \nbse ~the exceptional real space resolution of STM and the ability of the corresponding FT to identify and unambiguously differentiate between the two IC modes were successfully exploited to gain better insight into the formation and behaviour of the CDWs.

Analyses of large-scale LT STM images confirm that both modes indeed likewise modulate the type-I and type-III columns, while the type-II columns remain unmodulated. It is also shown that the amplitudes of both modes depend strongly on the bias voltages used in STM experiments. In images recorded under conditions that show both modes with similar amplitudes "beating" between them is clearly detected.

The suggested temperature dependent instabilities with the formation of CDW nanodomains would not result in a detectable difference in diffraction experiments in comparison to sliding of CDWs along distinct columns  On the other hand, STM offers an unique view, as it enables studying the surface with atomic resolution while still allowing discrimination between both CDW modes. Although it is not possible to unambiguously confirm the nanodomain model,  the data presented in this paper strongly supports it. The possibility of easy and frequent interchanging of \q and \qq offers an alternative mechanism of CDW sliding which provides an additional argument why only certain quasi-1D compounds exhibit CDW sliding.

\section*{Acknowledgements}
The work was supported by the Slovenian Research Agency (ARRS); Core Research Funding No.P1-0099 (E.Z.,  H.J.P.v.M., M.A.v.M., A.P.), young researcher project No. PR-07586 (M.A.v.M.), the CO Nanocenter and by the Natural Science and Engineering Research Council of Canada (J.C.B.).

\bibliographystyle{apsrev4-1} 
\bibliography{./bib/references}

\end{document}